\documentclass[journal,onecolumn]{IEEEtran}
\usepackage{ifpdf}
\usepackage{cite}
\usepackage{amsmath}
\usepackage{array}
\usepackage{amssymb}

\usepackage{url}

\usepackage{booktabs} 
\usepackage[numbers]{natbib}
\usepackage{algpseudocode}
\usepackage{algorithm}
\usepackage{empheq}
\usepackage{graphicx}
\usepackage{subcaption}
\usepackage{comment}
\usepackage{xcolor}
\begin{document}


\title{Enforcing Optimal Moving Target Defense Policies}
\title{Markov Decision Process to Enforce Moving Target Defence Policies}


\author{\IEEEauthorblockN{Jianjun Zheng and Akbar Siami Namin} \\
\IEEEauthorblockA{Computer Science Department\\
Texas Tech University\\
Lubbock, Texas, USA \\
Email: jianjun.zheng@ttu.edu, akbar.namin@ttu.edu}
}

\maketitle

\begin{abstract}

This paper introduces an approach based on control theory to model, analyze and select optimal security policies for Moving Target Defense (MTD) deployment strategies. A Markov Decision Process (MDP) scheme is presented to model states of the system from attacking point of view. The employed value iteration method is based on the Bellman optimality equation for optimal policy selection for each state defined in the system. The model is then utilized to analyze the impact of various costs on the optimal policy. The MDP model is then applied to two case studies to evaluate the performance of the model.
\end{abstract}


\IEEEpeerreviewmaketitle

\section{Introduction}
The essence of Moving Target Defense is security through diversification, in which the configurations and properties of a target system are dynamically and randomly changed \citep{DBLP:journals/jcst/ZhengN19}. This creates a complex and unpredictable moving target while making it computationally expensive for attackers to exploit the target system. While it increases the attack cost and degrades  the attackers' incentives, moving target defense can also impose some cost on the defenders and thus on the network infrastructure. Therefore, it is important to incorporate cost in the computational factors that may affect the effectiveness of MTD with respect to the type of attacks, environment, deployment, and employed MTD strategies \citep{Skowyra2016, Zheng2016, Maleki2016, Prakash:20153, Wright:2016}.

Another limiting factor of the implementation of MTD in practice is due to the security policies that are often defined across the network, on which the prospective MTD system would be deployed. These security policies not only regulate actions that are allowed or prohibited under certain circumstances but might also cause some conflicting issues with actions permitted by the MTD implementation.


Game theoretical approaches model the interactions between defenders and attackers, as the players of a game, and thus adopt the strategies usually employed by players. In such games, each player tries to determine the optimal strategy in order to maximize their incentives. Assuming that all players (defenders and attackers) in the game are rational, they tend to choose the best possible strategies to maximize their expected payoffs while minimizing their costs at each move. When the game reaches a state called ``{\it Nash Equilibrium},'' at which no player could increase their payoffs by changing strategies, the solution to the game at this state is considered to be optimal. However, in cyber defense the assumption that attackers would make rational decisions at each move to maximize their payoffs might not hold completely, mainly because attackers would try unpredictable actions to breach the system.

A Markov model is a stochastic model used to describe the state transition of a system. When combined with game theory, a Markov game model can describe the interaction between defenders and attackers and thus it would be possible to analyze the possible outcome of the system when it is in a certain state. The Markov chain game model is descriptive and useful for defenders by which necessary information is provided to them in order to choose the best strategy for the next move. However, the key challenge is that network defenders in some situations may not have access to the needed time to make informed decisions with respect to the feedback received from a model. As a result, a model is preferable that can make decisions to enforce proper security policies (i.e., actions) in certain circumestances.

To meet this challenge, this paper proposes to use Markov Decision Process (MDP) to model the state transition of a system in which the interaction between defenders and attackers is modeled through transitions from one state to another. The model incorporates the costs of players' actions and the existing security policies in a system using Bellman Optimality Equations in order to identify the optimal defense strategies or policies under different scenarios. The model enables the defender to analyze the impact on the policy change by the cost of strategy. This paper completes our preliminary work on modeling MTD using MDP \cite{Zheng2018} as follows:

\vspace{0.1cm}
\begin{itemize}
    \item Presents a more comprehensive Markov model to embrace additional dynamic nature of networks,
    \item Investigates the impact of different costs on the selection of optimal policy,
    \item Evaluates the proposed model through two use cases to demonstrate the applicability and usefulness of the model.
\end{itemize}

\vspace{0.15cm}

The remainder of the paper is organized as follows: 
Section \ref{sec:model} describes Markov Decision Process game model and Bellman Optimality Equation. Section \ref{sec:simulation} presents the model simulation setup, simulation results, and implications of findings. Section \ref{sec:conclusions} concludes the paper and sketches the future research directions.

\section{A Markov Decision Process-based Model}
\label{sec:model}
The interaction between a defender and an attacker is abstracted out as a discrete, finite-state, and finite-action Markov Decision Process (MDP). The model is formulated as a 4-tuple $(S, A, P, R)$, where:

\begin{itemize}
\renewcommand\labelitemi{--}
\item $S$ is the finite set of states.
\item $A$ is the finite set of control actions.
\item $P$ is the probability of transition from one state to another upon performing an action.
\item $R$ is the expected immediate rewards received after state transition associated with the control action performed.
\end{itemize}
    
\begin{figure}[t!]
  \begin{subfigure}[b]{0.5\textwidth}
    \includegraphics[width=\textwidth]{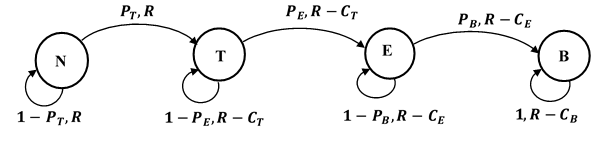}
    \caption{when the ``{\it Wait}'' action is taken.}
    \label{fig:transition_wait}
  \end{subfigure}
  \hfill 

  \begin{subfigure}[b]{0.5\textwidth}
    \includegraphics[width=\textwidth]{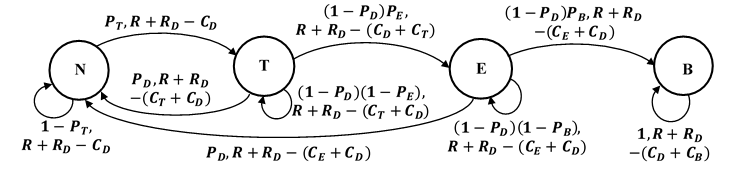}
    \caption{when the ``{\it Defend}'' action is taken.}
    \label{fig:transition_defende}
  \end{subfigure}
    \hfill 

  \begin{subfigure}[b]{0.47\textwidth}
    \includegraphics[width=\textwidth]{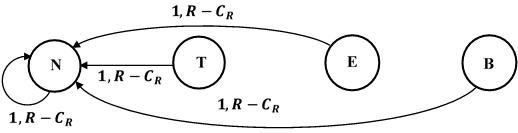}
    \caption{when the ``{\it Reset}'' action is taken.}
    \label{fig:transition_rest}
  \end{subfigure}
  \caption{State transition probabilities/costs of the MDP model.}
\end{figure}

\begin{figure*}
\includegraphics[height=2in, width=\textwidth]{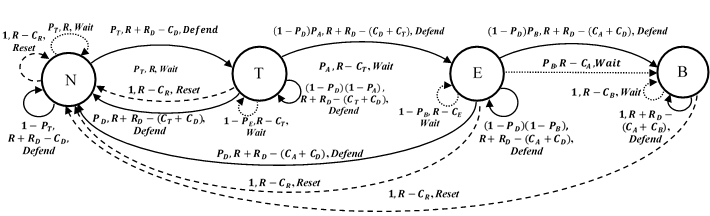}
\caption{The complete MDP model with state transition probabilities and costs \cite{Zheng2018}.}
\label{fig:model}
\end{figure*}

Figures \ref{fig:transition_wait}--\ref{fig:transition_rest} illustrate the state transition probabilities and costs under each control action. Furthermore, Figure \ref{fig:model} depicts the big picture of the proposed MDP-based model by combining Figures \ref{fig:transition_wait}--\ref{fig:transition_rest}. In the proposed model, the security defense mechanism is abstracted out into four states $(S)$ and three control actions $(A)$, as follows: 

\vspace{-0.2cm}
\begin{equation}
  S \in \begin{cases}
    N &  \textit{System Running Normally} \\
    T &  \textit{System Being Targeted} \\
    E &  \textit{System Being Exploited} \\
    B &  \textit{System Being Breached}
  \end{cases}
\end{equation}

\vspace{-0.15cm}
\begin{equation}
  A \in \left\{ \textit{Wait}, \textit{Defend}, \textit{Reset}\right\} \\  
\end{equation}

\vspace{-0.35cm}

\subsection{State Transitions}
Two scenarios are presented to help better understand the state transition in the proposed model:

\textit{Scenario 1} : Under the ``{\it Wait}''  control action, state {\itshape N} may transition to state \emph{T} with the probability values of $ P_T $. It may also return to itself with the self-transition probability of $ 1-P_T $ (i.e., the sum of all probabilities transitioning out of a state under a control action must be equal to $1$).

\textit{Scenario 2}: Under the ``{\it Wait}'' control action, a transition from  state \textit{T} to state \textit{E} occurs with probability $ P_E $ and the self-transition of  state \textit{T} occurs with probability $ 1-P_E $. Both transitions incurs a cost denoted by $ C_T $. Therefore, the immediate reward for each transition is $ R-C_T $.

In an analogous way, similar scenarios can be developed for the ``{\it Defend}'' and ``{\it Reset}'' control actions.

\subsection{Key Concepts of MDP}

In a typical MDP, the most critical property that must be satisfied is known as the \textit{Markov property}. This property states that the effects of an action taken in any state depend only on that state and not on the prior history or knowledge. 

A {\it policy} $ \pi $ in MDP is a mapping function from states to actions: $ \pi: S\rightarrow A $. In other words, a policy dictates each process (i.e., agent) to take certain actions while being in a specific  state.

The \textit{value function}, denoted by $V^\pi(s)$, represents the expected value of the received rewards, starting from state $S=s$ and following policy $\pi$. It is also called {\it state value function} or {\it utility function} and it is computed through the following equation:

\vspace{-0.45cm}
\begin{equation} 
\begin{split}
V^\pi(s_t) 
 & = \mathbb{E^\pi}[R_{t+1} + \gamma R_{t+2} + \gamma^2 R_{t+3}  \cdots |S=s_t] \\
 & =  \mathbb{E}^\pi[R_{t+1}+\gamma(R_{t+2}+\gamma R_{t+3} +  \cdots)|S=s_t] \\
  & = \mathbb{E}^\pi[R_{t+1} + \gamma V^\pi(s_{t+1})|S=s_t]
\end{split}
\end{equation}

where:
\begin{itemize}
\renewcommand\labelitemi{--}
\item $s_t$ denotes the state $S$ at the time interval $t$.
\item $\mathbb{E}^\pi$ is the expected value of total rewards gained by following policy $\pi$.
\item $R_{t+1}$ is the reward gained at state $S$ at $t+1$.
\item $\gamma$ is the discount factor.
\item $V^\pi(s_t)$ denotes the value function in the state $S$ at $t$.
\end{itemize}

By omitting the subscript of the time interval, the general form of the utility function can be represented as follows:

\vspace{-0.35cm}
\begin{align} \label{equ:utilizity}
V^\pi(s) &= \sum_{s'\in S}P(s,\pi,s')[R(s,\pi,s')+\gamma V^\pi(s')] 
\end{align}
\vspace{--0.25cm}
where:
\begin{itemize}
\renewcommand\labelitemi{--}
\item $P(s,\pi,s')$ is the transition probability starting from state $s$ and ending at state $s'$ after following policy $ \pi $.
\item $ R(s,\pi,s') $ is the expected rewards received after state transition from $ s $ to $ s' $ after following policy $ \pi $.
\item $ \gamma $ is the discount factor.
\end{itemize}

The \textit{discount factor} in MDP, denoted by $\gamma \in (0,1)$, presents the portion of the future rewards that would be lost in comparison to the present rewards. A smaller $ \gamma $ means the rewards received in the future would worth much less than the present rewards due to the discount. As a result, the agent should follow the policy and collect the rewards immediately instead of waiting to collect and claim them in the future.

Finally, an \textit{optimal policy} $\pi^{*}$ is a
control action $a \in A$ that generates the maximum state value function and is expressed by {\it Bellman Optimality Equation} \citep{Bellman2010}:

\vspace{-0.35cm}
\begin{equation}
V_{i+1}^*(s)= \max_{a\in A}\sum_{s'\in S}P(s,a,s')[R(s,a,s')+\gamma V_i^*(s')]
\end{equation}

where
 $V_{i+1}^*(s)$ is the value function in the state $S$ by following the optimal policy. The optimal policy can be obtained by solving the MDP problem or the Bellman Optimality Equation.

\subsection{Solving MDP}
\label{subsec:solving}

The MDP problem can be solved by using the value iteration method developed by Bellman \citep{Bellman2010}. The value iteration method is simple and intuitive.  The algorithm of value iteration is listed in Algorithm \ref{alg:value_iteration}.

\begin{algorithm}
\caption{Value Iteration.}
\begin{algorithmic}[1]
\State initialize $ V_0(s)=0 $, $\forall s \in S, \epsilon=  $ a small positive number.
\State $ \Delta \gets 0, i \gets 0 $
\Repeat
\State  For each {$ s \in S $}
\State\hspace{0.5em} $ V_{i+1}(s) \gets \max_{a \in A}P(s,a,s')[R(s,a,s')+\gamma V_i(s')] $
\State\hspace{0.5em} $ \Delta \leftarrow \max(\Delta, |V_{i+1}(s)-V_i(s)|)  $
\State $i \gets i+1$
\Until {$ \Delta < \epsilon $}
\State Output the policy $ \pi $, such that $$ \pi(s) \gets arg\max_{a \in A}\sum_{s' \in S}P(s,a,s')[R(s,a,s')+\gamma V(s')]  $$
\end{algorithmic}
\label{alg:value_iteration}
\end{algorithm}

The following example demonstrates how to solve the Bellman Optimality Equation using Algorithm \ref{alg:value_iteration}.

As shown in Figure \ref{fig:model}, there are three control actions $a \in \{Wait, Defend, Reset\}$ at each state. To find the optimal policy at a specific state, say state $ E $, we use Equation  \ref{equ:utilizity} to calculate the state value under each of the three control actions and then assign the maximum of the three values computed for the three actions to the state value. The process will be repeated until $V_{i+1}^*(E)$ converges, (i.e. $V_{i+1}^*(E) \approx V_i^*(E)$). At the convergence, the control action corresponding to the maximum state value will be selected as the optimal policy. 
\vspace{-0.15cm}
\begin{equation*}
 V_{i+1}^*(E)=\max_{a \in A}\sum_{s' \in S}P(E,a,s')[R(E,a,s')+\gamma V_i^*(s')] \medskip 
\end{equation*}

\vspace{-0.38cm}
\begin{equation*}
\scalebox{0.92}{%
$  \gets \max \left \{
\begin{array}{ll}
P(E,a,E)[R(E,a,E)+\gamma V_i^*(E)] \\
\mbox{ }+P(E,a,B)[R(E,a,B)+\gamma V_i^*(B)] & a=\mbox{Wait} \\ \mbox{} \\
P(E,a,N)[R(E,a,N)+\gamma V_i^*(N)] \\ 
\mbox{ }+P(E,a,E)[R(E,a,E)+\gamma V_i^*(E)] \\
\mbox{ }+P(E,a,B)[R(E,a,B)+\gamma V_i^*(B)] & a=\mbox{Defend} \\ \mbox{} \\
P(E,a,N)[R(E,a,N)+\gamma V_i^*(N)] & a=\mbox{Reset}
\end{array} 
\right.\bigskip
$}
\end{equation*}

After plugging in all variables shown in Figure \ref{fig:model}, the Bellman Optimality Equation at state $E$ in our model can be written as:
\begin{equation*}
\scalebox{0.95}{%
$ \gets \max \left \{
\begin{array}{ll}
(1-P_B)[(R-C_A)+\gamma V_i^*(E)] \\ 
\mbox{ } + P_B[(R-C_A)+\gamma V_i^*(B)] & a=\mbox{Wait} \\
\mbox{} \\
P_D[R+R_D-C_A-C_D)+ \\ \mbox{} \gamma V_i^*(N)] 
+(1-P_D)(1-P_B) \\ \mbox{} \*[(R+R_D-C_A-C_D) +\gamma V_i^*(E)] \\
\mbox{ } + (1-P_D)P_B[(R+R_D \\ -C_A-C_D) + \gamma V_i^*(B)] & a=\mbox{Defend} \\
\mbox{} \\
(R-C_R) + \gamma V_i^*(N) & a=\mbox{Reset}
\end{array}
\right. \bigskip
$}
\end{equation*}

Similarly, we can derive the equations to calculate  $V_{i+1}^*(N)$, \\ $V_{i+1}^*(T)$, and $V_{i+1}^*(B)$.

\subsection{Cost Impact on Optimal Policy}
The reward that is gained by taking a specific action has a significant impact on the calculation of the optimal policy. Hence, a defender can control the optimal policy by changing the reward. In our model, we introduce the cost factor and define the expected reward as the result of the baseline reward $ R $ subtracted by the cost incurred by an action ``\textit{a}'' during a state transition. The action can be initiated by an attacker or the defender. For example, an MTD-based action from the defender can incur the cost $C_D$, an exploitation action from the attacker can incur the cost $C_E$, and a reset action can incur the cost $C_R$. After plugging these costs into the cost factor, the Bellman equation will be:

\vspace{-0.46cm}
\begin{equation}
\begin{split}
V_{i+1}^*(s)=\max_{a \in A} \sum_{s' \in S}P(s,a,s')[(R-C(s,a,s'))  & \\ + \gamma V_i^*(s')] 
\end{split}
\end{equation}
This equation will allow us to analyze the cost impact on the optimal policy. Several other control theoretic-based problems can be formulated using MDP (e.g., \citep{TavakoliDC16, DBLP:conf/bigdataconf/ZhengN18, TavakoliDC19}) in a similar manner.

\section{Simulation and Results}
\label{sec:simulation}
This section presents the results of a simulation with the goal of analyzing the optimal policy in various network attack dynamics and then evaluating the impact of the cost on deciding about the optimal policy.

\subsection{Network Environment}
The initial assumption is that our network is infrequently targeted or exploited by attackers. Therefore, we set the corresponding probability values very low (e.g., $P_T=0.2, P_A=0.2$). Furthermore, we assume that an MTD defense technique is actively applied to protect the network: (e.g., $ P_D=0.6$), and the probability of system breach is medium: (e.g., $P_B=0.4$). We also assume the baseline reward of the system is $R=10$ and the reward received by performing the defense action is $R_D=5$. Moreover, the discount factor is fixed and it is set to $\gamma = 0.9$.

\subsection{Defense: Cost vs. Optimal Policy}
To investigate the impact of defense cost on the optimal policy, we fix the other costs as follows: 

\begin{itemize}
\renewcommand\labelitemi{--}
\item $C_T = 0.1$: the cost incurred by attacker's reconnaissance (i.e., system being targeted),
\item $C_E=3$: the cost incurred by attacker's exploitation,
\item $C_B=4$: the cost incurred by system breach, and 
\item $C_R=4$: the cost incurred by resetting the system.
\end{itemize}

\begin{figure}[t!]
  \begin{subfigure}[b]{0.45\textwidth}
    \includegraphics[width=\textwidth]{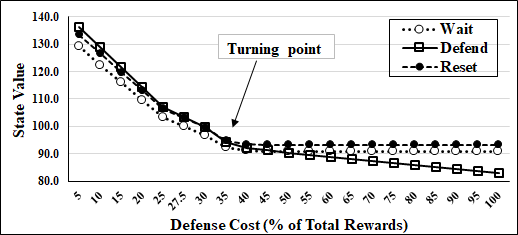}
    \caption{Impact of defense cost on state value at state $S=E$.}
    \label{fig:vf_defense}
  \end{subfigure}
  \hfill 
  
  \begin{subfigure}[b]{0.45\textwidth}
    \includegraphics[width=\textwidth]{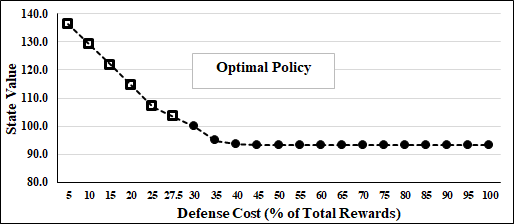}
    \caption{Optimal policies as the defense cost increases.}
    \label{fig:optimal}
  \end{subfigure}
  \hfill 
  \caption{Defense impact and optimal policy.}
\end{figure}

We implemented Algorithm \ref{alg:value_iteration} with $\epsilon=0.001$ using Microsoft Excel and calculated the value function for each policy (wait, defend, reset) at each state $S \in \{N,T,E,B\}$ with different values for the defense costs. Figure \ref{fig:vf_defense} shows the state value vs. the defense cost at state $S=E$.

Figure \ref{fig:vf_defense} demonstrates a clear trend in which the state value decreases as the defense cost increases. However, when the defense cost exceeds a certain point (i.e., 35\% of the total rewards), the state values corresponding to the Wait and Reset actions stay unchanged; whereas, the state value corresponding to the ``Defend" action continues to decrease. This observation implies that the defense cost does not have impact on these two policies. This behavior can be explained by analyzing Equation \ref{equ:utilizity}. Since each value function consists of the immediate reward $ R $ and the discounted value of the successor state $\gamma V(s')$, we can re-write Equation \ref{equ:utilizity} in a more general form as follows:

\vspace{0.2cm}
\scalebox{0.95}{%
$ V_{i+1}^*(E)=\max \left \{
\begin{array}{ll}
R_1 + \gamma[V_i^*(E)+V_i^*(B)] & a=\mbox{Wait} \\
(R_2 -kC_D) + \gamma[V_i^*(N) \\ 
\mbox{ } + V_i^*(E) + V_i^*(B)] & a=\mbox{Defend} \\ 
R_3 + \gamma V_i^*(N) & a=\mbox{Reset}
\end{array}
\right.
$}

Where:
\begin{itemize}
\renewcommand\labelitemi{--}
\item $R_1, R_2, R_3$ are the sum of the immediate rewards of respective value function.
\item $k$ is the sum of all coefficients of $C_D$ in value function for the ``Defend'' action.
\end{itemize}

For example, when $a=Wait$, 
$V_{i+1}^*(E)= R_1 + \gamma[V_i^*(E)+V_i^*(B)]$. In this equation, the immediate reward $R_1$ is independent of $C_D$, but $ \gamma[V_i^*(E) + V_i^*(B)] $ is related to $ C_D$ when $C_D$ is small. However, when $C_D$ exceeds a certain value $C_{wait}^*$,  $\gamma[V_i^* (E)+V_i^* (B)]$ becomes a constant and therefore independent of $C_D$. This pattern holds for the value functions when $a=Defend$ and $a=Reset$. As a result, the above equation can be further written in terms of $C_D$:
\vspace{-0.15cm}
\setcounter{equation}{3}
\begin{subequations}\label{eq:4}
\begin{gather}
V_{i+1,Wait}^*(E)=\begin{cases} R_1+\beta_1C_D & C_D < C_{wait}^* \\ c & C_D \geq C_{wait}^* \end{cases} \\
V_{i+1,Reset}^*(E)=\begin{cases} R_2+\beta_2C_D & C_D < C_{defend}^* \\ c & C_D \geq C_{defend}^* \end{cases} \\
V_{i+1,Defend}^*(E)=\begin{cases} (R_3 - kC_D)+\beta_3C_D & C_D < C_{reset}^* \\ R_3-kC_D & C_D \geq C_{reset}^* \end{cases}
\end{gather}
\end{subequations}

With respect to Equations $(4a)$ and $(4b)$, we can reason that when $C_D$ is small, both value functions are linear with respect to $C_D$, but with different slopes. In an analogous way, when $C_D$ is large, both value functions remain constant. The value function for the ``Defend'' action, however, is always a linear function of $C_D$, but the slope changes when $C_D$ exceeds a certain value, as shown in $(4c)$.

A similar pattern is observable through Figure \ref{fig:vf_defense}. It shows that when the defense cost is below a certain value, called the {\it turning point}, the ``Defend' action is the optimal policy. But when the defense cost is above the turning point, the ``Reset'' action turns out to be the optimal policy because it generates higher rewards than the other two actions. This optimal policy shift can be better shown when we plot the optimal state value at each level of the defense cost for state $S=E$ in Figure \ref{fig:optimal}. The first 6 data points represent  the ``Defend'' action and the remaining data points show the ``Reset'' action.

Another interesting finding in Figure \ref{fig:vf_defense} is that when the defense cost exceeds a certain value (45\% in the simulation experiment) and if the optimal policy (the ``Reset'' action) is not available, then the ``Wait'' action is a better option than the ``Defend'' action. It is because the cost of defending the system would be higher than the damage caused by the attacker. Under this situation, the defender could choose to wait and accept the damage until system reset is available again.

\begin{table}[H]
\caption{Optimal policies as the defense cost increases.}
\begin{center}
\begin{tabular}{lcccc}
\toprule
Cost (\%) & $S=N$ & $S=T$ & {\bf S=E} & $S=B$ \\
\midrule
5&Defend&Defend&\textbf{Defend}&Reset \\
10&Defend&Defend&\textbf{Defend}&Reset \\
... & ... & ... & \textbf{...} & ... \\
25&Defend&Defend&\textbf{Defend}&Reset \\
27.5&Defend&Defend&\textbf{Defend}&Reset \\
30&Defend&Defend&\textbf{Reset}&Reset \\
35&Wait&Defend&\textbf{Reset}&Reset \\
40&Wait&Defend&\textbf{Reset}&Reset \\
... & ... & ... & \textbf{...} & ... \\
95&Wait&Wait&\textbf{Reset}&Reset \\
100&Wait&Wait&\textbf{Reset}&Reset \\
\bottomrule
\end{tabular}
\end{center}
\label{tab:optimalchanges}
\end{table}

To help the defender choose the optimal policy at a given state, we calculate all optimal policies for all four states with various defense costs. The results are listed in Table \ref{tab:optimalchanges}. At state $S=E$ (marked in bold), for instance, when the defense cost is less than or equal to 27.5\% of the total rewards, the best action to take (a.k.a the optimal policy) is "Defend". However, when the defense cost is above 27.5\% of the total rewards, the best action to take will change to "Reset." The table confirms the optimal policy shift behavior as the defense cost increases.

\subsection{Reset: Cost vs. Optimal Policy}

In this simulation, we analyze the impact of the reset action cost on the optimal policy. With respect to previous simulation, the parameters remain unchanged, except $C_D$, which is set to $C_D=4$. We executed the simulation using the same procedure as described in Section \ref{subsec:solving}. 

\begin{figure}[t!]
  \begin{subfigure}[b]{0.45\textwidth}
    \includegraphics[width=\textwidth]{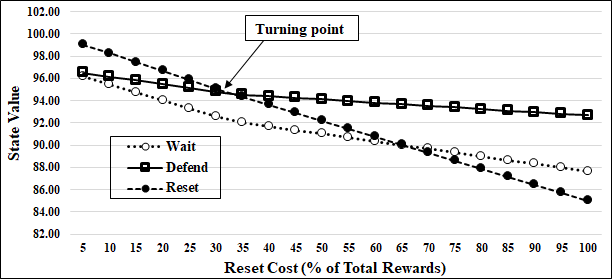}
    \caption{Impact of reset cost on state value at state $S=E$.}
    \label{fig:resetcost}
  \end{subfigure}
  \hfill 
  \begin{subfigure}[b]{0.45\textwidth}
    \includegraphics[width=\textwidth]{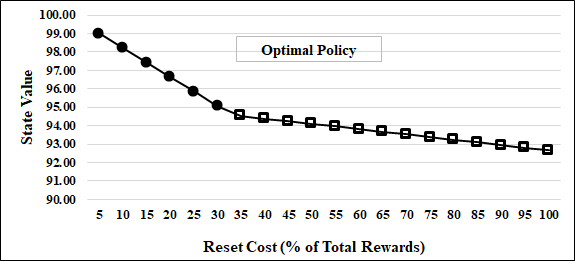}
    \caption{Optimal policy changes as the reset cost increases.}
    \label{fig:optimalreset}
  \end{subfigure}
  \caption{Reset impact and optimal policy.}
\end{figure}

The results are plotted in Figures \ref{fig:resetcost} and \ref{fig:optimalreset}. As shown in Figure \ref{fig:resetcost}, when the cost of reset action increases, all state values decrease implying that the reset cost has impact on all three policies. In other words, the greatest impact on the policy will be achieved if the underlying system is reset. Conversely, it has the least impact on the policy if the system is in defence mode. Unlike Figure \ref{fig:vf_defense} for ``defense", the impact for reset cost is not stabilized. Employing similar technique presented in Section \ref{subsec:solving}, this pattern can be explained by rewriting Equation \ref{equ:utilizity} in terms of the reset cost $C_R$. 

Equations $(6a)$, $(6b)$, and $(6c)$ show that all value functions are linear in terms of $C_R$ with different slopes. The slopes change once $C_R$ exceeds a certain value.

Figure \ref{fig:resetcost} also shows a similar optimal policy shift as in the previous simulation (i.e., defense) in that when the reset cost is below the turning point, the ``Reset'' action is the optimal policy. However, when the reset cost is above the turning point, the ``Defend'' action turns out to be the optimal policy. Figure \ref{fig:optimalreset} is plotted to demonstrate this optimal policy shift. The first 6 data points represent the ``Reset'' policy; whereas, the remaining data points represent the ``Defend'' policy.

\vspace{-0.3cm}
\setcounter{equation}{5}
\begin{subequations}
\begin{gather}
V_{i+1,Wait}^*(E)=\begin{cases} R_4+\beta_4C_R & C_R < C_R^1 \\ R_4+\beta_5C_R & C_R \geq C_R^1 \end{cases} \\
V_{i+1,Reset}^*(E)=\begin{cases} R_5+\beta_6C_R & C_R < C_R^2 \\ R_5+\beta_6C_R & C_R \geq C_R^2 \end{cases}  \\
V_{i+1,Defend}^*(E)=\begin{cases} (R_6-C_R)+\beta_7C_R & C_R < C_R^3 \\ (R_6-C_R)+\beta_8C_R & C_R \geq C_R^3 \end{cases}
\end{gather}
\end{subequations}

A similar finding in this simulation is that when the reset cost exceeds a certain value (i.e., 65\% in this experiment) and if the optimal policy (the ``Defend'' action) is not available somehow, then the ``Wait'' action is better than the ``Reset'' action. It is because the cost of resetting the system is much higher than the damage caused by the attacker. In other words, if for any reason a defense action is not available, then it would be better to just accept and absorb the damage (i.e., do nothing) than resetting the system.

\subsection{Exploitation: Cost vs. Optimal Policy}
Similar to the previous simulations, the parameters remain unchanged, except $C_E$, which is set to $C_E=4$. We analyze the impact of the cost incurred by attacker's exploitation on the optimal policy. We run this simulation using the same procedure as described in Section \ref{subsec:solving}.

\begin{figure}[t!]
  \begin{subfigure}[b]{0.45\textwidth}
    \includegraphics[width=\textwidth]{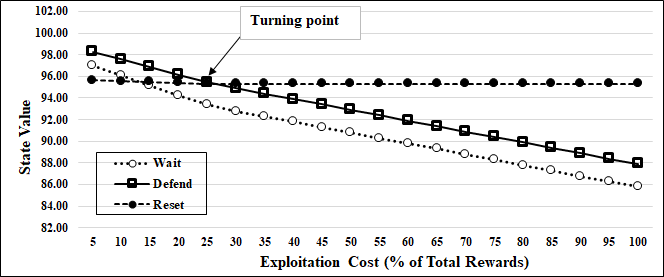}
    \caption{Impact of exploitation cost on state value at state $S=E$.}
    \label{fig:impactexplo}
  \end{subfigure}
  \hfill 
  \begin{subfigure}[b]{0.45\textwidth}
    \includegraphics[width=\textwidth]{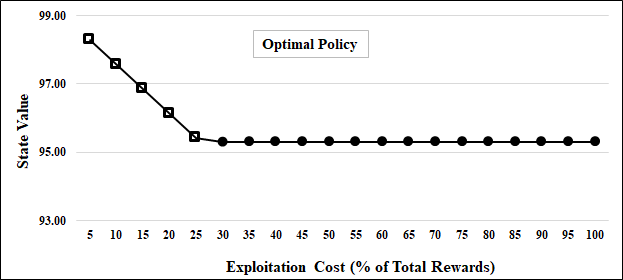}
    \caption{Optimal policy changes as the exploitation cost increases.}
    \label{fig:optimalexplo}
  \end{subfigure}
  \caption{Exploitation Impact and optimal policy.}
\end{figure}

The results are plotted in Figure \ref{fig:impactexplo} and \ref{fig:optimalexplo}. These figures exhibit similar trends as observed through the previous simulations. We also plot the optimal policy with respect to the defense and exploitation costs to show how the optimal policy shifts in different network environment. 

\begin{figure}[t!]
  \begin{subfigure}[b]{0.22\textwidth}
    \includegraphics[width=\textwidth]{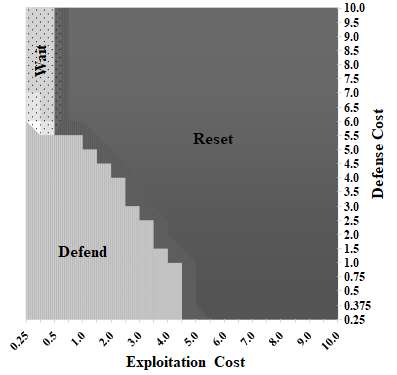}
    \caption{With respect to the defense cost and the exploitation cost.}
    \label{fig:optimaldefenseexplo}
  \end{subfigure}
  \hfill 
  \begin{subfigure}[b]{0.22\textwidth}
    \includegraphics[width=\textwidth]{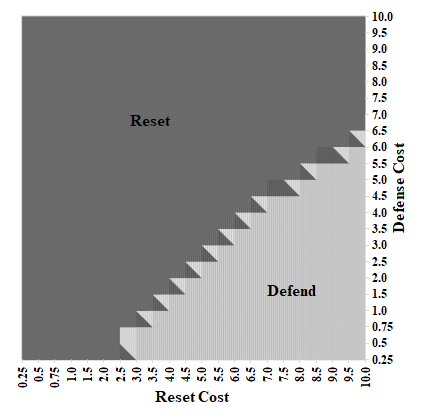}
    \caption{With respect to the defense cost and the reset cost.}
    \label{fig:optimaldefensereset}
  \end{subfigure}
  \caption{Optimal policy.}
\end{figure}

Figure \ref{fig:optimaldefenseexplo} visualizes the optimal selection of policies. For example, if a combination of the defense and exploitation costs falls within the area on the top layer (i.e., high defense and high exploitation costs), then ``Reset'' will be the optimal policy. On the other hand, if the combination falls within the left bottom corner (marked in blue: high defense  but low exploitation costs), then ``Wait'' will be the optimal policy. Finally, if the combination falls within the triangle-shape area on the middle layer, then ``Defend'' will the optimal policy.

\subsection{Case Studies}
We demonstrate how to apply our model to two existing Moving Target Defense techniques: 1) Decoy-based MTD, and 2) Self Cleansing Intrusion Tolerance (SCIT). It is important to note that the goal is to qualitatively analyze the feasibility of these two techniques and not their effectiveness.

\subsubsection{Decoy-based MTD}

A decoy system is a phony platform with appealing appearance but fake data and credentials that is developed for the purpose of trapping unauthorized users. These types of systems are usually used to study the attacker's exploitation patterns in order to protect real targets. Decoy-based MTD is one of the newest ideas, in which  decoy hosts are mixed with the protected target hosts on the network. Decoy hosts may confuse attackers by providing fake data and thus they slow down the pace of attacks towards the real target host. Meantime, the defender can gain additional valuable time to prepare for defending against the next attack. 

To achieve this goal, first, the configurations of each decoy system need to be similar to the target host so that it becomes hard to distinguish the decoy system from the main operational one. Second, the ratio of the target hosts to the decoys must be very low, for example $1:10,000$ as recommended by the NCLYS 2009 \cite{NCLYS2009}. However, research shows that the ratio of $1:99$ \citep{Skowyra2016} is still effective for some attacks. 

It is reasonable to assume each host on the network to employ a similar defense mechanisms so that each host looks legitimate and indistinguishable. Furthermore, the ratio of the target hosts to the decoys is kept at least $1:10,000$. With respect to these two assumptions, we observe that:

\begin{enumerate}
\item The total defense cost is the sum of each host's defense cost. As a result, the defense cost of a large-scale decoy-based MTD can be very high.
\item When the configuration of the target host is changed, all decoys must be reset and cloned to match the new configuration. Therefore, the reset cost of the decoy-based MTD can also be very high.
\item Each decoy is loaded with fake data and accounts. Hence, the cost of a decoy breach will be very low if not negligible. For the same reason, the exploitation cost is also very low.
\end{enumerate}

\begin{figure}[t!]
  \begin{subfigure}[b]{0.22\textwidth}
    \includegraphics[width=\textwidth]{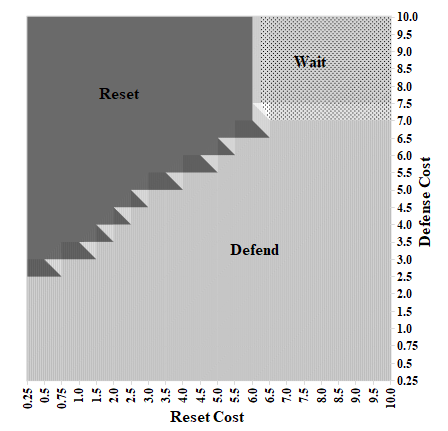}
    \caption{Decoy-based MTD.}
    \label{fig:decoy}
  \end{subfigure}
  \hfill 
  \begin{subfigure}[b]{0.22\textwidth}
    \includegraphics[width=\textwidth]{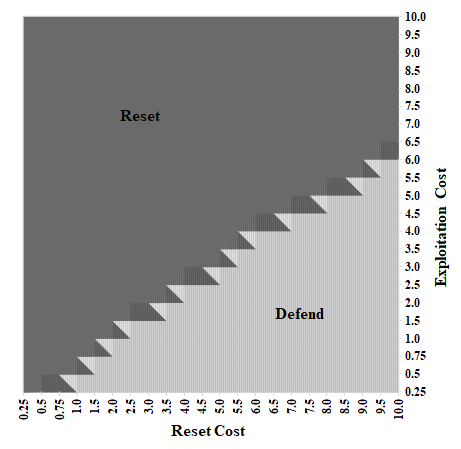}
    \caption{SCIT-based MTD.}
    \label{fig:SCIT}
  \end{subfigure}
  \caption{Optimal policies in case studies.}
\end{figure}

After plugging in the parameters in our simulation, we plot the optimal policy with respect to the defense and rest costs as shown in Figure \ref{fig:decoy}. We can observe that when both the defense and reset costs are high, the optimal policy is ``Wait.'' In reality, however, this may not be acceptable. Therefore, unless there is some way to reduce the defense cost or the rest cost, our model shows decoy-based MTD might not be a viable defense technique.

\subsubsection{Self Cleansing Intrusion Tolerance (SCIT)}
SCIT \citep{Bangalore2009} periodically rotates an array of virtual machines to reduce the exposure time of the protected server with the aim of disrupting potential attacks. 
 According to the literature, the rotation frequency is within minutes and VM reset happens at each rotation, incurring a reset cost. The reset cost is not discussed as a factor for the effectiveness of the technique. Therefore, we performed a simulation similar to the one reported in this paper and plotted the results in Figure \ref{fig:SCIT}.

As Figure \ref{fig:SCIT} indicates, only when either the reset cost is low or the exploitation damage (cost) is high, the optimal policy is ``Reset.'' For example, in a mission-critical infrastructure where the potential exploitation damage might be very high, it is preferable to periodically reset the system to disrupt potential attacks, then SCIT might be a viable option. On the other hand, if with new technology, the cost of resetting and reloading a VM can be reduced, then SCIT might be adopted.

\section{CONCLUSIONS AND FUTURE WORK}
\label{sec:conclusions}
In this paper, we proposed to use finite-state, finite-action, and stationary Markov Decision Process to model the interaction between a defender and an attacker. We investigated three possible defense strategies (wait, defend, and reset) for each of the four system states and used Bellman Optimality Equation to mathematically define the optimal policy. By solving the Bellman Optimality Equation, we were able to find the optimal policy when the system was in a specific state. Various simulation experiments were conducted to demonstrate how the optimal policy would shift when cost change. The optimal policy shift can help defender to make right decision at various situation. We also applied our model in two existing MTD techniques and case studies to evaluate the feasibility of the two MTD techniques.

We plan to include some quantitative cost estimation methods in our model to fine tune the cost analysis of our research. We also plan to extend our model to Partially Observable Markov Decision Process (POMDP) and compare the accuracy of the two models. Another research opportunity is to utilize the big data research to estimate the transition probabilities of different control actions in our model in a real-world dataset using techniques such as deep learning algorithms \citep{DBLP:conf/icmla/Siami-NaminiTN18, simaarchive} and techniques based on evidence theory \citep{DBLP:conf/bigdataconf/ChatterjeeND18, Moi2018}. Also, formal techniques can be used in conjunction with stochastic approaches to model the MTD problem 
\citep{DBLP:conf/compsac/SartoliN16a, DBLP:conf/sac/SartoliN17}.

\section*{Acknowledgement}
\label{sec:acknowledgement}
This project is funded in part by grants (Awards No: 1516636 and 1564293) from National Science Foundation.

\bibliographystyle{IEEEtran}

\bibliography{IEEEabrv,IEEEexample}

\end{document}